\documentclass[aps,pr,groupedaddress,floatfix]{revtex4}
 
\usepackage{graphicx}
\usepackage{subfig}
 
\usepackage{amsmath,amssymb,amsfonts}
\usepackage{mathtools}
 
\newcommand\be{\begin{equation}}
\newcommand\ee{\end{equation}}
\newcommand\bea{\begin{eqnarray}}
\newcommand\eea{\end{eqnarray}}

\begin{document}

\title{Incorporating the Hayflick Limit \\ into a model of Telomere Dynamics}


\author{Benoit M. Cyrenne}
\affiliation{Department of Physics, Engineering Physics, and Astronomy, Queen's University, Kingston, ON K7L 3N6, Canada}
\author{R. J. Gooding}
\affiliation{Department of Physics, Engineering Physics, and Astronomy, Queen's University, Kingston, ON K7L 3N6, Canada}
 
\date{\today}

\begin{abstract}

A model of telomere dynamics is proposed and examined. Our model, which extends a previously introduced two-compartment model that incorporates stem cells as progenitors of new cells,  imposes the
Hayflick Limit, the maximum number of cell divisions that are possible.  This new model leads to  cell populations for which the average telomere length is not necessarily
a monotonically decreasing function of time, in contrast to previously published models. We provide a phase diagram indicating where such results would be expected. In addition, qualitatively
different results are obtained for the evolution of the total cell population. Last, in comparison to available  leukocyte baboon data, this new model is shown to provide a better fit to 
biological data. \\
\end{abstract}


\maketitle

\section{Introduction}

Telomeres were first recognized in the 1930's as special structures capping the ends of chromosomes, the cell's genetic material. Further research has identified telomeres to be important components in a variety of cellular processes, the most important being replication \cite{blackburn91}.  Cellular replication, known as mitosis, involves one cell dividing itself into two identical copies, each containing a full copy of the genetic material. 

The replication of the cell's linear chromosomes is a inherently destructive process, in that each replication causes a shortening of the chromosome \cite{olovnikov73}.  Cells are able to protect their (coding) genetic material by instead allowing telomeres to shorten \cite{blackburn91}.  As telomeres have a finite length, this puts an upper bound, called the Hayflick Limit, on the number of divisions a cell is able to undergo \cite{hayflick61,shay00}.  The decreasing lengths of telomeres have been found to be involved in aging, with shorter telomeres associated with advanced age, and the mortality of an organism associated with the finite length of the telomere.

While all cells contain chromosones and telomeres, there are only a select few cell types that are actively dividing throughout adulthood. Most other cells instead have lifetimes on the same order as the organism itself, and thus divide rarely or not at all. Leukocytes, or white blood cells, are among the few cell types that regularly divide. Leukocytes encompass multiple different types of mononucleated blood cells including lymphocytes, \textit{viz.} B cells and T cells, as well as granulocytes. These cells are involved in the cellular component of immune response, and thus they experience large cellular turnover and are required to maintain large, dynamic  populations. The ease with which leukocytes can be extracted and analyzed, especially when compared to other rapidly dividing cell types, such as colon crypt cells, makes them the perfect subject for analyzing the Hayflick limit and telomere length dynamics.

All blood cells, white or red, originate as hematopoietic stem cells within the bone marrow. These stem cells are pluripotent, self-renewing cells capable of differentiating into unique types of hematopoietic progenitor cells, which then mature into peripheral blood cells \cite{sidorov09}. The hematopoietic stem cells are considered to be ``immortal cells", and are believed to maintain their telomere length by means of the enzyme telomerase. The kinetics of T cells and other leukocytes has been studied extensively, and values have been obtained for their proliferation and the death rates.  Estimated values are shown in Table 1.  

\begin{table}[h!]
\centering
\begin{tabular}{| l || c |  c |}
\hline
{\bf Cell Type} & {\bf Mitotic Rate (/week)} & {\bf Death Rate (/week)} \\ \hline\hline
Memory T-Cells  \cite{borghans07}& 0.02 - 0.03   & 0.02 - 0.03   \\ \hline
Mature B-Cells  \cite{cooperman04} & 0.5   & 0.48 \\ \hline
Neutrophiles \cite{summers10}  & ~0.042  & ~0.042  \\ \hline
\end{tabular}
\caption{Estimated values from the literature for the mitotic and death rates for (memory) T cells, (memory) B cells and (neutrophiles) granulocytes.}
\end{table}

As cells divide and multiply the end-replication problem causes chromosomal shortening.  The level of attrition in leukocytes has been measured and most estimates of the number of base pairs lost per division range from 50 to 100 bp, with the highest estimate at 120 bp/division \cite{weng95,vaziri93}.  This value is not uniform across species or even between individuals, with the exact figure changing greatly between individuals and different cells types.  As telomeres are of a finite length, the chromosonal loss that each mitotic cycle causes results in the Hayflick limit, the theoretical maximum number of divisions a cell may undergo.  Leukocytes  were found to be able to divide on average 23 times before reaching senescence \cite{pilyugin97}.  Similiar to the amount of base pairs lost per cell division, the Hayflick limit varies greatly among different cell types. For human fibroblasts between 14 and 29 divisions were observed, while for human embryos the limit was found to be between 35 and 63 divisions \cite{hayflick61}. 

\subsection{Outline of Paper}

In the next section we state and put into context previously published dynamical models of telomere shortening. Using for now a minimalist's approach, we restrict our attention to the simplest
linear dynamical models, as our goal is to report on the dynamics that results from the incorporation of the Hayflick Limit into such minimalist models. We compare our results to previously published work
and find one new feature -- a phase transition occurs in terms of the qualitative behaviour of the mean telomere length. Then we discuss the application of our model to experimental data
on baboon leukocytes, and show that one obtains an improved fit to this data using our model. Finally, we discuss how this model can be related to aging and cancer and other aspects of
chromosomal length dynamics.

%\newpage
 
\section{Summary of Previous Dynamical Models}
\label{sec:previouswork}

We briefly review previously studies models below, allowing us to define our notation, and to allow for a direct comparison of our work with previously published theories.

\subsection{Single compartment model}

Our introduction highlights that the majority of situations in which dividing cells are encountered, such as leukocytes, involve a stem cell population
replenishing a population of (white blood) cells. If one ignores the stem cells, one may model the simpler situation of a population of cells dividing
and dying \cite{deboer98}.

Let $i(t)$ denote the number of divisions that a cell has undergone at time $t$, and let $N_i(t)$ denote the number of such cells in the population. If
the rate of mitotic division is denoted by $M$, and the rate of cell death by $D$, allowing for potentially an \textit{infinite} number of cell divisions, one simply
has
\bea\label{eq:1comp}
\dot N_0~&=&~-(M+D)~N_0\nonumber\\
\dot N_i~&=&~2M~N_{i-1}~-(M+D)~N_i~~~~~~~i=1,2,\dots
\eea
Assuming that
\be\label{eq:initN}
N_i(0)~=~N_o^o~\delta_{i,0}
\ee
the total number of cells in this model, denoted by $\mathcal N(t)=\sum_i N_i(t)$, is easily found to be
\be\label{eq:singlecompN}
\mathcal N(t)~=~N_o^o~e^{(M-D)t}
\ee
If $D>M$ the total number of cell decreases with time, as expected physically. However, for $D>M$ the total cell population diverges in this model. Note that
Table 1  does not make clear which inequality is to be expected, and we discuss this result below.  

The average number of cell divisions undergone in such a cell population is found through the use of the generating function \cite{berg78}
\be\label{eq:GenFunc}
\mathcal F(x,t)~=~\sum_i ~x^i~N_i(t)
\ee
whose ODE (derivative with respect to time) is easily solved using the original equations of motion.
That is (using the same initial conditions),
\be\label{eq:singlecompibar}
\overline i (t)~=~\frac{1}{\mathcal N (t)}~\sum_j ~j ~N_j(t)~=~\frac{1}{\mathcal N (t)}~\frac{\partial \mathcal F}{\partial x}\vert_{x=1}~=~2Mt
\ee

Let $L(t)$ denote the length of a chromosome at time $t$. Define the initial length of a chromosome's telomere to be $L_o$, and suppose that, on average, 
the number of base pairs lost from the telomeres in each mitotic event is given by $\Delta_o$. Then
\bea\label{eq:defineLbar}
\overline L (t)~&\equiv&\frac{1}{\mathcal N(t)}\Big( N_0 L_o~+~N_1(L_o-\Delta_o)~+~N_2(L_o-2\Delta_o)~+~\dots \Big)\nonumber\\
&=&~L_o~-~\Delta_o~\overline i (t)
\eea
giving that for this one-compartment model
\be\label{eq:singlecompLbar}
\overline L (t)~=~L_o~-~\Delta_o~2Mt
\ee
This shows that, on average, in this simple single compartment model the length of a telomere decreases linearly with time. This generates the unphysical result
that Eq.~(\ref{eq:singlecompLbar}) predicts that eventually the length of a telomere is negative.

Below we contrast our results with these results, in particular Eqs.~(\ref{eq:singlecompN},\ref{eq:singlecompLbar}).

\subsection{Two compartment model without the Hayflick Limit}

The work of Itzkovitz \textit{et~al.} \cite{PRE2008} advanced that of the previous subsection by including a second compartment, the latter associated
with a population of stem cells that replenish the population of cells under study. The modification of the above model in one sense is very minor,
namely the equations of motion now become
\bea\label{eq:2comp}
\dot N_0~&=&~\alpha-(M+D)~N_0\nonumber\\
\dot N_i~&=&~2M~N_{i-1}~-(M+D)~N_i~~~~~~~i=1,2,\dots
\eea
where $\alpha$ corresponds to the rate at which the stem cells are producing new cells. It is assumed that all nuclei produced by stem cells have
telomeres of length $L_o$. However, a more substantive change was made in that $\alpha$ is chosen to maintain a constant cell population size
(independent of the distribution of telomere lengths), this condition corresponding to  
\be\label{eq:contantN}
\alpha~=~(D-M)~N_o^o
\ee
(This result is for the same initial condition as given in Eq.~(\ref{eq:initN}); more generally, a constant total cell population is found for $\alpha~=~(D-M)~N_o$~,
where $N_o$ is the total population of all cells at $t=0$.)
As shown in that paper, again utilizing the generating function of Eq.~(\ref{eq:GenFunc}), one then obtains that the average length of the telomeres follows
\be
\label{eq:2compLbar}
\overline L(t)~=~L_o~-~\Delta_o \frac{2M N_o^o}{\alpha}~\big(1-e^{-(\alpha~ t/N_o^o)}\big)
\ee
We see that while this two-compartment model includes the unphysical situation of having telomeres with a negative length, the result for the average telomere length
now remains positive when
\be\label{eq:posLbar}
L_o~>~\Delta_o \frac{2M N_o^o}{\alpha}
\ee
Moreover, for small times one obtains
\be
\overline L(t\approx 0)~=~L_o~-~\Delta_o ~2Mt
\ee
which is the same result as is found for the one-compartment model in Eq.~(\ref{eq:singlecompLbar}).

We have also analysed this model \textit{without} the imposition of a fixed population size -- that is, for which Eq.~(\ref{eq:contantN}) is not satisfied. Using the initial condition of Eq.~(\ref{eq:initN}), one obtains
\be
\mathcal N (t)~=~N_o^o~e^{(M-D)t}~+~\frac{\alpha}{M-D}~\big(e^{(M-D)t}~-~1\big)
\ee
which is the similar to the one-compartment model -- see Eq.~(\ref{eq:singlecompN}) -- with the added effect of the stem cell pool. Note again that for $M>D$, like the one-compartment
model, this expression diverges for large times. In addition, one can derive the expression for the average
telomere length. This expression is cumbersome, but its behaviour can be identified for small and long times, namely
\[
 \overline i (t) \sim \begin{dcases*}
        ~~2Mt & for $t\approx 0^+$\\
        ~~2Mt & for $t\rightarrow\infty$ and $M>D$\\
        ~~\frac{2M}{D-M}& for $t\rightarrow\infty$ and $D>M$
        \end{dcases*}
\]
Note the similarity of this results, when $M>D$, to the one-compartment model result of Eq.~(\ref{eq:singlecompibar}).
Most importantly, one may show that $\overline i (t)$ for the two-compartment model without a fixed total population is a monotonically increasing function of time, regardless of the sign of $M-D$.

%\newpage

\section{Incorporating the Hayflick Limit in a Two-Compartment Model}
\label{sec:formalism}

\subsection{Dynamical System of Equations}
\label{subsec:HFEOMs}

We modify the two-compartment model as follows. We assume that after mitosis has occurred a fixed number of times, the Hayflick limit, which we denote by $h$, the cells
become senescent. For simplicity, we assume that the death rate of the senescent cells is the same as that for any other cells, although it is simple to relax this assumption. The
resulting dynamical system of equations is therefore
\bea\label{eq:HFlimitEOM}
\dot N_0~&=&~\alpha-(M+D)~N_0\nonumber\\
\dot N_i~&=&~2M~N_{i-1}~-(M+D)~N_i~~~~~~~i=1,2,\dots,h-1\\
\dot N_h~&=&~2M~N_{h-1}~-~D~N_h~~\nonumber
\eea
As discussed in the introduction, the biologically relevant values for $h$ are between 14 and 63, and for human leukocytes a value of $h=23$ has been measured.
This system of equations is easily solved for any $h$, simply by iteration. However, it is easiest to work with a scaled system of equations, as we now discuss.

One can reduce Eq.~(\ref{eq:HFlimitEOM}) to a sequence of dimensionless rate equations, and similarly dimensionless parameters.  To do so we redefine the generational populations, 
$N_i$ in terms of a new set of  scaled populations, $f_i$, as follows.  Let
\bea
N_i ~&=&~a_i~f_i~~~~~~i=0,1,2,\dots,h~~~~~~~\\
t~&=&~b~\tau\nonumber
\eea
and choosing
\be\label{eq:scalefactors}
a_i = \Big(\frac{2M}{M+D}\Big)^i\Big(\frac{\alpha}{M+D}\Big)~~~~~i = 0,1,2,\dots,h~~~~~~~b = (M+D)^{-1}
\ee 
the system of equation governing the generational populations then reduce to the much simpler form
\bea\label{eq:scaledEOMs}
\frac{d}{d\tau}f_0(\tau) &=& 1 - f_0(\tau)\nonumber\\
\frac{d}{d\tau}f_I(\tau) &=& f_{I-1}(\tau) - f_I(\tau)~~~~~~~~~~I=1,2,\dots,h-1\\
\frac{d}{d\tau}f_h(\tau) &=& f_{h-1}(\tau) - \gamma~f_h(\tau) \nonumber
\eea
where
\be
\gamma \equiv \frac{D}{D+M}
\ee
Using the initial condition of Eq.~(\ref{eq:initN}) we see that solution of Eq.~(\ref{eq:scaledEOMs}) depends on only three dimensionless parameters, namely
\be
f_o^o~\equiv~\frac{M+D}{\alpha}~N_o^o
\ee
$\gamma$ and, of course, $h$. Below we investigate the dynamics of this model subject to variations in these quantities.

For completeness, we note that steady-state solutions of these equations are then trivially seen to be
\be
f_I (\infty)=1~~~~~~~~I=0,1,\dots,h-1,~~~~~~~~~f_h (\infty)=\frac{1}{\gamma}
\ee
Clearly, the introduction of the scale factors of Eq.~(\ref{eq:scalefactors})  dramatically simplifies the considerations of the statics of this problem. (Nonlinear aspects of the steady-state solutions
found when the rate constants are modelled with Hill functions, etc, were considered by others \cite{pilyugin97}.)

\subsection{Phase Diagram of Dynamical Solutions}
\label{subsec:phasediagram}

As mentioned above, one may solve the system of equations iteratively. However, as is apparent from Eq.~(\ref{eq:scaledEOMs}) it is simplest to integrate the scaled equations. Expressions
for $f_I$ for $I=0,\dots,h-1$ can be written down, namely
\bea\label{eq:sols}
f_o(\tau)&=& e^{-\tau} \big(f_o^o-1+e^{\tau})\nonumber\\
f_1(\tau)&=& e^{-\tau} \big(f_o^o~t-1-t+e^{\tau})\nonumber\\
f_2(\tau)&=& e^{-\tau} \big(f_o^o~t^2-2t-t^2+2e^{\tau})\\
f_3(\tau)&=&\dots\nonumber
\eea
The expression for $f_h(\tau)$ is long and complicated (even for $h=5$ -- see below) so we refrain from providing it. A formal
closed-form solution for related equations may be obtained, if desired \cite{gyllenberg86}. (Analytical expressions
found from the integration of these coupled ODE's using Laplace transforms, or a computer algebra software, such as \textit{Mathematica}$^\copyright$, can be obtained in a straightforward manner.)
Here we show how two qualitatively different classes of solutions exist, one of which is not found in either the one compartment model, or the two-compartment model without the Hayflick limit.

\begin{figure}[t]
\centering
\parbox{6.7cm}{
\includegraphics[width=6.2cm]{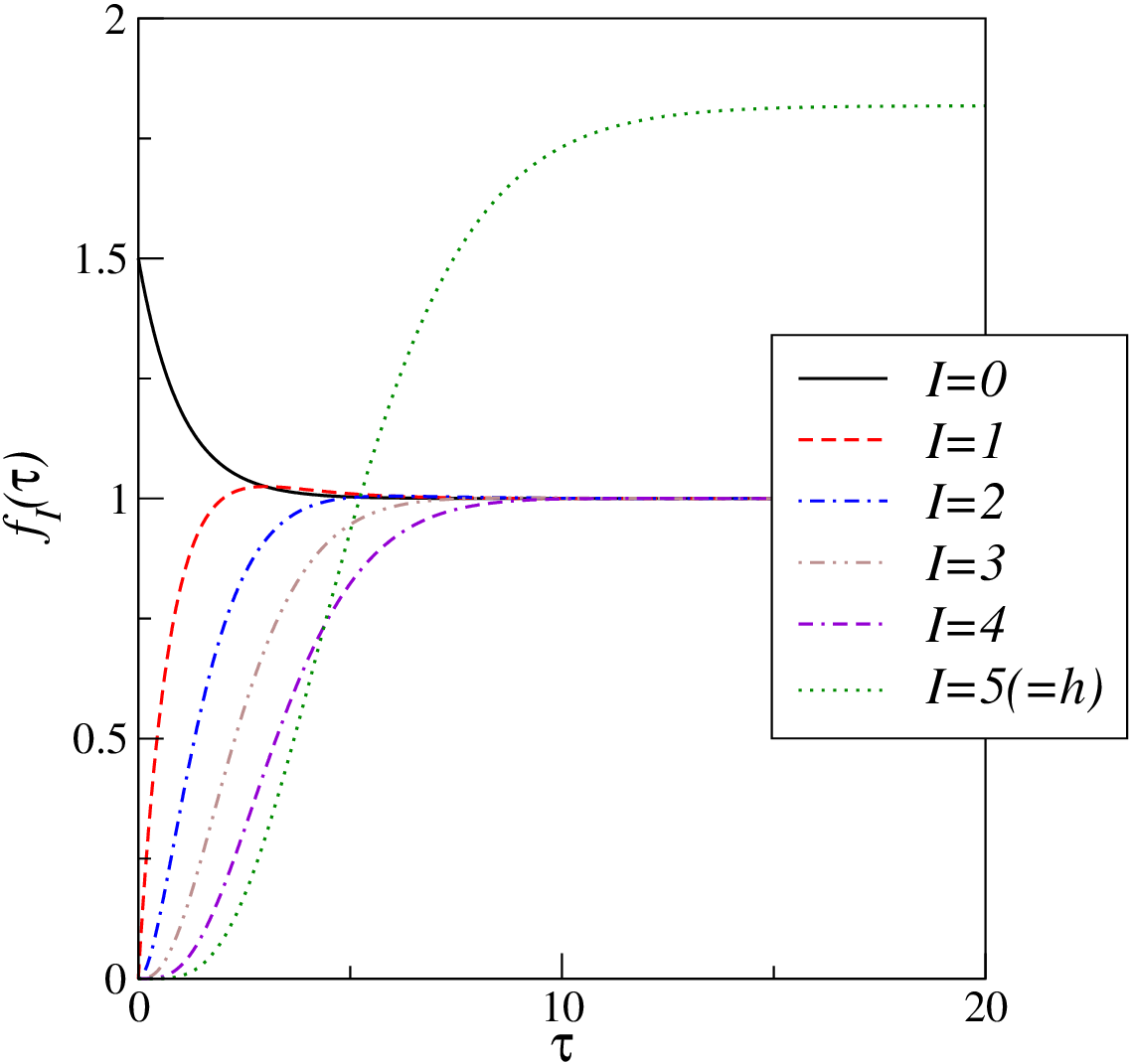}
\label{fig:2figsA}}
\qquad
\begin{minipage}{6.5cm}
\includegraphics[width=5.5cm]{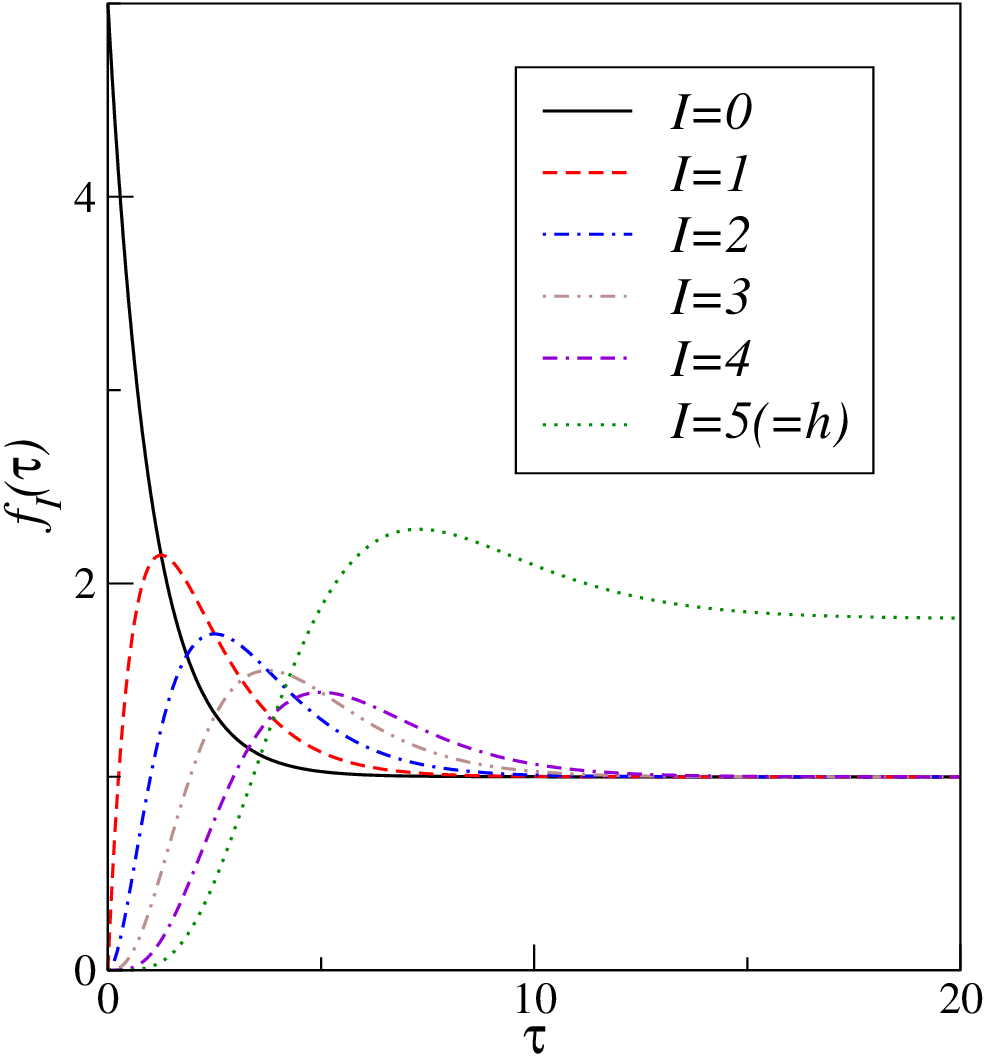}
\end{minipage}
 \caption{\label{fig:2scaledfs} The  behaviour of the (scaled) populations, $f_I(\tau)$, $I=0,1,2,3,4,5$, for $\gamma=0.55$ for (a) $f_o^o=1.5$ and (b) $f_o^o=5$. In the first
figure one sees that the populations (for $I>0$) increase monotonically with $\tau$, whereas in the second figure  (for $I>0$) the populations achieve a local maximum.}
\end{figure}

To be concrete, consider $h=5$; similar behaviour is found for larger $h$, but here we choose this small value to make the presentation of our results simpler. It is clear
that on biological grounds we expect $\gamma\approx 1/2$ since $D\approx M$. Therefore, we show the solution of $f_I(\tau)$ for $I=0,1,...,5$ for $\gamma=0.55$ for two values of $f_o^o$, namely
$f_o^o=1.5$ and 5. Our results are shown in Fig.~\ref{fig:2scaledfs} (note that qualitatively similar results are found when $\gamma=0.45$, corresponding to $D<M$).
For the smaller value of $f_o^o$ the (scaled) populations for $I=1,2,\dots,h$ all increase monotonically with time, whereas for the larger value of $f_o^o$ the populations all show a
local maximum (located at different $\tau$ for each $f_I$). The appearance of these local maxima could be important in the interpretation of biological experiments, as we now describe. 

\begin{figure}[t]
%\begin{center}
\includegraphics[scale=0.45]{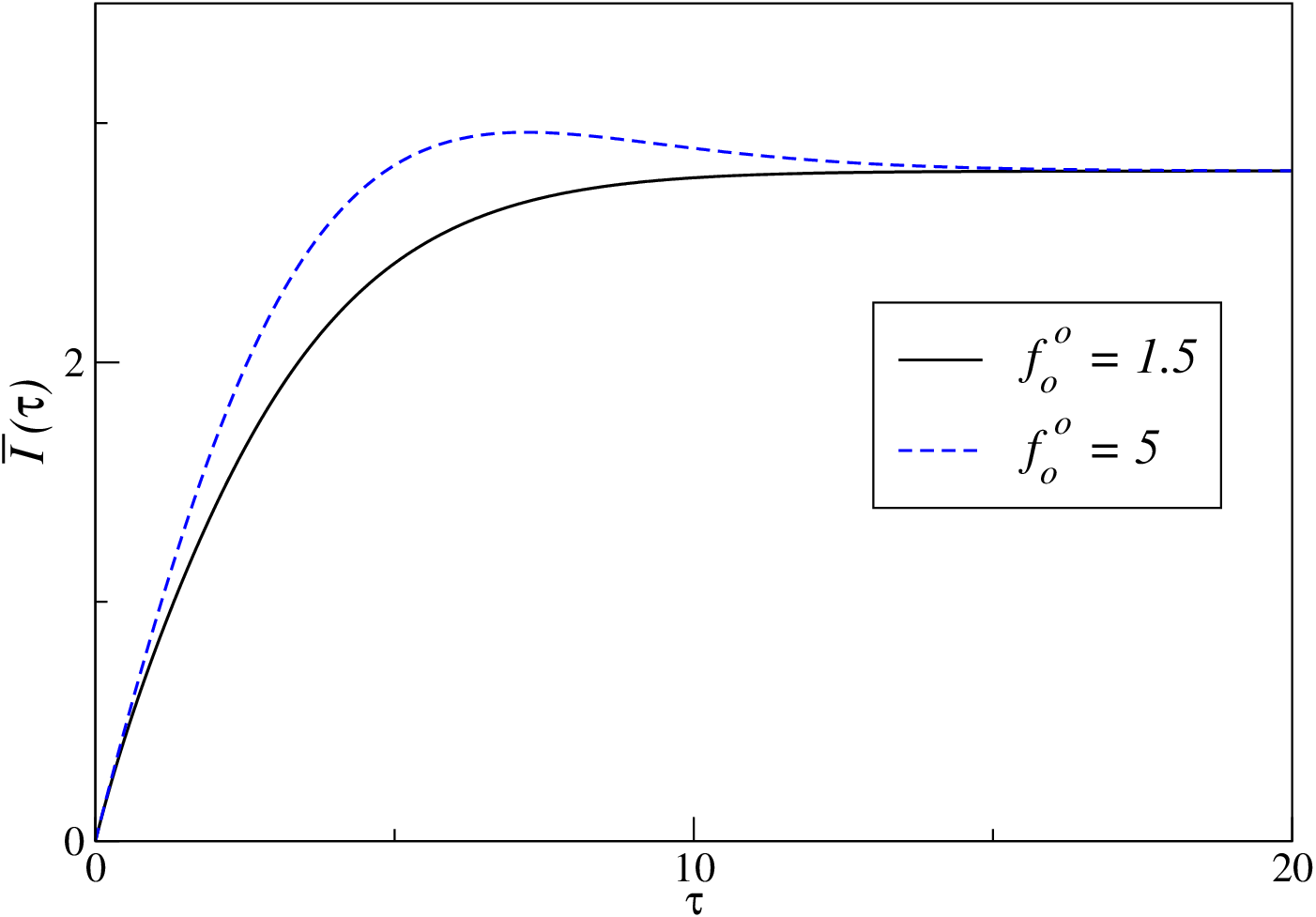}
%\end{center}
\caption{\label{fig:bumps} The variation in the average number of cell divisions in the (scaled) population of cells - see Eq.~(\protect\ref{eq:scaledIbar}). The maximum number
of divisions is $h=5$, we chose $\gamma=0.55$, and the curves are for small $f_o^o(=1.5)$ (solid line) and large $f_o^o(=5)$ (dashed line). Similar results are found for $\gamma=0.45$,
but as discussed in the text, the appearance or absence of a local maximum in $\overline I (\tau)$ is dependent on the choice of $h$, $\gamma$ and $f_o^o$.}
%\end{center}
\end{figure}

These model equations were motivated by the behaviour of telomere lengths as individuals age. Therefore, it is important to consider the average number of cell divisions that
are found for these (scaled) populations. This latter quantity is given by  
\be\label{eq:scaledIbar}
\overline I (\tau)~=~\frac{\sum_{I=0}^h~I~f_I(\tau)}{\sum_{I=0}^h~f_I(\tau)}
\ee
As seen in Fig.~\ref{fig:bumps}, for small $f_o^o$ this function increases monotonically with time, whereas for larger $f_o^o$ one obtains a local maximum at an intermediate time. 
The importance of this result is seen when considered with the definition of the average telomere length $\overline L (t)$ -- see Eq.~(\ref{eq:defineLbar}). That is, one can now obtain a local minimum for this quantity, and this could lead to significant differences when fitting (\textit{e.g.}, leukocyte) telomere data, as we describe in the next section.  (We have confirmed this behaviour of $\overline L (t)$ for a variety of different parameter values.)

The above result raises an interesting question: How does the appearance of a local maximum  in $\overline I (t)$ (or a local minimum in $\overline L (t)$) depend on the value of the
Hayflick limit, $h$? It's clear that in the $h\rightarrow\infty$ limit one recovers the two-compartment model discussed above, and when a constant total population size is enforced via
the imposition of Eq.~(\ref{eq:contantN}), from Eq.~(\ref{eq:2compLbar}) one sees that this
produces a monotonically decreasing $\overline L (t)$ \textit{without} any local maximum. Therefore, our simple linear model incorporating the Hayflick limit possesses a transition in that for finite $h$ 
and fixed $\gamma$, at small $f_o^o$ no
extrema appear in the dynamics, whereas for larger $f_o^o$ extrema do appear. One can therefore identify critical parameters $(\gamma,f_o^o)$ for each $h$
separating these qualitatively different dynamics. It is a simple manner to do this numerically.

\begin{figure}[t]
%\begin{center}
\includegraphics[scale=0.45]{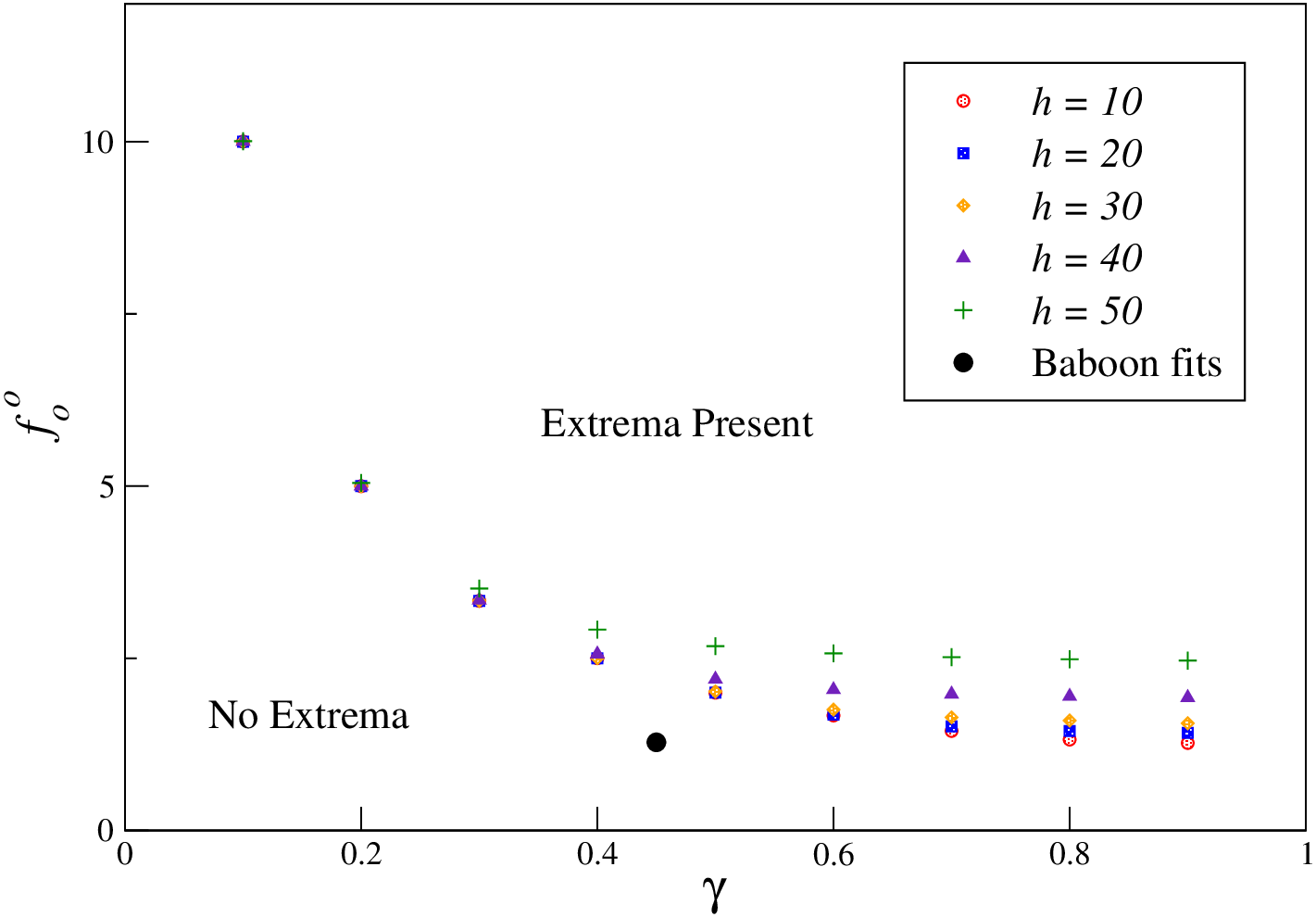}
%\end{center}
\caption{\label{fig:phasediag} A phase diagram showing how the average (scaled) telomere length of Eq.~(\protect\ref{eq:scaledIbar}) 
depends on the the maximum number of cell divisions, $h$ (the Hayflick Limit), $\gamma$, and $f_o^o$. Above the phase boundary
for each $h$, $\overline I (t)$ has a local minimum, whereas below the phase boundary no extrema are present. As $h$ increases
the region that includes an extremum in $\overline I (t)$ diminishes, and eventually this region vanishes as $h\rightarrow\infty$. The large solid
circle shows the average values of these parameters when fitted to baboon leukocyte data -- see \S \protect\ref{sec:data} for a discussion of this result.}
%\end{center}
\end{figure}

We have completed this analysis for $h=10,20,30,40,50$, and our results are shown in Fig.~\ref{fig:phasediag}. As expected based on the above-studied two-compartment model (see section 2.1 of this paper),  as $h$ increases the parameter regime over which one finds dynamics possessing extrema is found to shrink, and eventually this region must vanish in the $h\rightarrow\infty$ limit. This analysis can also be completed for the unscaled populations $N_i(t)$ and a phase diagram found in terms of $\overline L (t)$; our numerical result  for such a calculation is that the resulting phase diagram is identical to that of Fig.~\ref{fig:phasediag}.

\newpage

\section{Application to Leukocyte Data}
\label{sec:data}

In Ref. \cite{PRE2008} the two-compartment model with fixed population size was compared to experimental data \cite{baerlocher07} for 3 different baboons (subjects 17891, 17927
and 18760) for 3 different leukocytes - B cells,  T cells, and granulocytes. (Note that we confirmed numerically that this model represents an improvement over the one-compartment model of Ref. \cite{deboer98}.) Therefore, we then compared our model, which incorporates the Hayflick limit, to this same data set. For our comparison we used $h=23$, based on Ref. \cite{pilyugin97}, although our results are not sensitive to small changes in $h$ (that is, $h=20$ to $h=26$ all
give very similar fits). We find that fitting this same data with our Hayflick Limit model leads to a small reduction (6\%) in the mean square differences of the data and fitted functions relative to Ref. \cite{PRE2008}.

For all three baboons and and all three leukocytes, our results for the fitted parameters correspond to the ``No Extrema" portion of our dynamical phase diagram. Averaging values of $\gamma$ and
$f_o^o$ over the three different baboons and the three different leukocytes gives 
\be
\overline f_o^o~=~1.29\pm0.16~~~~~~~~~~~~~\overline \gamma~=~0.46\pm0.04\nonumber
\ee
showing that very similar parameters fit all of the experimental data sets. These average values are included on Fig.~\ref{fig:phasediag}, showing that for this $h=23$ dynamical system indeed no 
extremal dynamics are encountered.

Other interesting biological information contained in these fits (rate constants in units of inverse weeks) is as follows.
\begin{itemize}
\item The initial lengths, $L_o$, of the telomeres are different for different baboons, but do not vary as much between different leukocytes.
\item The shortening of the telomeres per mitotic division, $\Delta_o$, strongly varies between leukocytes for a given baboon, as well
as showing strong variation between baboons.
\item In all cases
the fitted values of the death and mitotic division rates are similar in that $0.39 \lesssim \gamma \lesssim 0.54$, showing that $D$ is similar to but
somewhat smaller than $M$. This result is similar to the previously published work summarized in our Table 1.
\item The rates at which the stem cells replenish the leukocytes are remarkably
similar: $390 \lesssim \alpha \lesssim 396$.
\end{itemize}

\section{Summary and Discussion}
\label{sec:discussion}

The dynamical equations describing telomere shortening are certain to be complicated. We have modelled this biology in terms
of the number of cells in a population which have undergone a given number of cell divisions up to a maximum number of divisions, called the Hayflick Limit. We have
taken a minimalist's approach in using simple linear equations of motion. Our reason is nothing more than wanting to identify the different qualitative dynamics
that result from such a ``bare bones" model. Then, when new complications are added to these equations, such as the effects of telomerase, \textit{etc.},
new biological behaviours can hopefully more clearly be identified.

Our model is a modification of the work of two-compartment model of Ref. \cite{PRE2008}, which itself is an extension of the one-compartment model of Ref. \cite{deboer98}.
In a two-compartment model one accounts for the presence of stem cells as being the progenitors of new cells with chromosomes having full-length telomeres. Both of these
models, \textit{viz.} Eqs.~(\ref{eq:1comp},\ref{eq:2comp}), suffer from the unphysical populations of negative length telomeres that are produced, a byproduct of these models not accounting for the Hayflick limit. Our model
limits the total number of mitotic events that can occur to a set of chromosomes, and therefore accounts for the Hayflick limit. One may analytically solve the resulting set of
dynamical equations for the cell populations, and from this calculate, for example, the average length of telomeres as a function of time.

The interesting result that follows from our model is shown in Fig.~\ref{fig:bumps}. That is, our  model, given in Eq.~(\ref{eq:HFlimitEOM}), can produce
an average telomere length that decreases with time, reaches a local extremum, and then \textit{increases} with time until some steady-state value is reached. The biological
importance of such a result can be reasoned as follows. In an experimental study of telomere lengths, if data is found that shows such non-monotonic behaviour,
it would not necessarily be correct to attribute such data to new biology, such as the lengthening of telomeres by telomerase (or ALT, or other chromosome altering events). 
Instead, the simplest dynamical equations that incorporate senescence and the Hayflick limit are sufficient to give such a result. We have compared the resulting expression
for average telomere length to published data, and reasonable agreement is found, being, in fact, somewhat of a better fit than is found using the previously published
two-compartment model with a fixed population size. 

As the above discussion foreshadows, the work presented in this paper is but a first step at addressing the complicated biology associated with telomere shortening. Future
work will investigate biology such as that found in malignant cancerous cells, otherwise known as ``immortal cells", leading to uncontrolled growth of tumours. Noting
that early work on this problem \cite{kim94} reported that in  all tumour cells lines, and in 90\% of all primary tumour samples (from biopsies) telomerase is present, 
modelling the feedback of this enzyme on chromosome biology deserves further attention.

\section*{Acknowledgements}

We thank Peter Lansdorp for sharing his baboon telomere length data with us. This work was supported, in part, by Queen's University.

% \vspace{1cm}
%\textbf{References}
%\bibliographystyle{natbib}
\bibliography{../telomere_bibtex}

\end{document}